\documentclass{article}
\usepackage{graphicx}
\usepackage{dcolumn}
\usepackage{bm}

\def\be{\begin{equation}}
\def\ee{\end{equation}}
\def\ba{\begin{array}}
\def\ea{\end{array}}
\def\bea{\begin{eqnarray}}
\def\eea{\end{eqnarray}}

\begin{document}

\title {Mass dependence of onset of multifragmentation in low energy
heavy-ion collisions}

\author{Yogesh K. Vermani and Rajeev K. Puri
\footnote{rkpuri@pu.ac.in} \\
Department of Physics, Panjab University, \\
Chandigarh-160014, India.}

\maketitle

\begin{abstract}
Based on the quantum molecular dynamics (QMD) picture, we
calculate the energy and mass dependence of fragment production.
For the present study, we simulated the reactions of
$^{20}Ne+^{20}Ne$, $^{40}Ar+^{45}Sc$, $^{58}Ni+^{58}Ni$,
$^{86}Kr+^{93}Nb$,$^{129}Xe+^{124}Sn$ and $^{197}Au+^{197}Au$ at
central geometry. Our findings clearly show a linear mass
dependence for the peak center-of-mass energy at which the maximal
IMF production occurs. Such linear dependence for peak
center-of-mass energy on the system size has also been observed in
recent experimental studies. We also predict a similar behavior
for the multiplicities of different kinds of fragments.
Experiments are called for to verify this prediction.
\end{abstract}


\vspace{2pc} \noindent{\it Keywords}: heavy-ion collisions,
multifragmentation, quantum molecular dynamics (QMD) model.
\maketitle

\section{\label{intro}Introduction}

The central heavy-ion (HI) collisions have been reported to result
into a complete disassembly of nuclear matter at bombarding
energies above 100 MeV/nucleon. This disassembly of hot and dense
nuclear matter also commonly known as the onset of
multifragmentation, is found to occur when nuclear density drops
to less than half of its initial value \cite{leray}. In the low
density phase, the onset of multifragmentation is expected to
occur due to Coulomb instabilities. It is well known that the mean
multiplicity of intermediate mass fragments $\langle N_{IMF}
\rangle$ depends strongly on the bombarding energy as well as on
the impact parameter of the reaction. Leray {\it et al}.
\cite{leray} studied the fragment distribution for the reaction of
O+AgBr near the point of threshold multifragmentation and reported
the onset of multifragmentation around 150 MeV/nucleon. Earlier,
Peilert {\it et al}. \cite{peilert} have shown that the true
multifragmentation events were confined to central collisions
only. In their study, $^{93}Nb+^{93}Nb$ reaction was simulated at
incident energies between 30 and 200 MeV/nucleon. In central
collisions of $^{93}Nb+^{93}Nb$, one observes the maximal
multiplicity $\langle M_{C} (A>4) \rangle$ around 100 MeV/nucelon.
More recently, an extensive and exhaustive study by Puri {\it et
al}. \cite{rkp} reported the outcome of $^{40}Ca+^{40}Ca$ reaction
over incident energies between 20 and 1000 MeV/nucleon and over
the entire impact parameter range. This study indicated the
generation of events from incomplete fusion-fission to
multifragment emission and finally complete disassembly of the
nuclear matter. They observed a peak in the fragment production
around 60 MeV/nucleon in the central collisions. Interestingly
enough, a rise and fall in the multiplicity with impact parameter
was not observed for low incident energies (20-40 MeV/nucleon).
The existence of peak energy for maximal IMF emission was in
accordance with an earlier study by Peilert {\it et al.}
\cite{peilert} using Au nuclei. The multifragmentation, therefore,
exhibits a complex picture which is quite sensitive to the
entrance channel characteristics \emph{i.e.}, to the impact
parameter, beam energy as well as to the total mass of the target
and projectile \cite{leray, will, blaich, beau, li, li2}. The beam
energy dependence of IMF emission was recently analyzed by Sisan
{\it et al.} \cite{sis} using MSU 4$\pi $-Array set up. In their
study, emission of intermediate mass fragments was reported for
the central collisions of $^{40}Ar+^{45}Sc$, $^{58}Ni+^{58}Ni$ and
$^{86}Kr+^{93}Nb$. They predicted a rise and fall in the emission
of IMFs with beam energy and observed a linear peak energy
dependence on the size of the system. The percolation calculations
used in the above study, however, could not fully explain this
dependence. This led to the conclusion by Sisan {\it et al.} that
perhaps phase space models can explain this observation
\cite{sis}. We plan to address this situation by employing a
dynamical model, where one can follow the reaction dynamics from
the start to the end where matter is cold and fragmented. Our
present study employs microscopic {\it quantum molecular dynamics}
(QMD) model \cite{aich, hart} which is described in section
~\ref{model}. Section ~\ref{results} is devoted to model
calculations and results, which are finally concluded in section
~\ref{summary}.

\section{\label{model}Description of the model}

The {\it quantum molecular dynamics} (QMD) model is a time
dependent A-body theory which is able to describe the many body
phenomenon like fragment formation. Here each nucleon in phase
space is represented by a Gaussian wave packet of the form:
\begin{equation}
{\psi}_i({\bf r},{\bf p}_i(t),{\bf r}_i(t))=\frac{1}{(2\pi
L)^{3/4}} exp \left[ \frac{i}{\hbar} {\bf p}_i(t)\cdot {\bf
r}-\frac{({\bf r}-{\bf r}_i(t))^2}{4L} \right]. \label{s1}
\end{equation}
Mean position $r_{i}(t)$ and mean momentum $p_{i}(t)$ are the two
time dependent parameters. The Gaussian width $\sqrt{L}$ is
centered around the mean position $r_{i}(t)$ and mean momentum
$p_{i}(t)$ and is same for all nucleons. This value of $\sqrt{L}$
corresponds to a root-mean-square radius of each nucleon. The
effect of different Gaussian width in fragmentation is reported in
reference \cite{jai2}. The centroids of Gaussian wave packets
$({\bf r}_i(t)$, ${\bf p}_i(t))$ in phase space follow the
Hamilton's equations of motion \cite{aich, hart}:
\begin{equation}
\dot{{\bf p}}_i=- \frac{\partial \langle H \rangle}{\partial {\bf
r}_i}; ~\ \dot{{\bf r}}_i=\frac{\partial \langle
H\rangle}{\partial {\bf p}_i}.
\end{equation}

In the above equations, $\langle H \rangle$ stands for the total
Hamiltonian of the system, which consists of kinetic and potential
energy terms:
\begin{equation}
\langle H \rangle= \sum _{i=1}^{A_{T} +A_{P}} \frac{{\bf
p}_{i}^{2} }{2m_{i}} + \frac{1}{2}\sum _{i;j \neq i}^{A_{T}
+A_{P}} V_{ij}^{loc} +V_{ij}^{Yuk}+ V_{ij}^{Coul}, \label{H}
\end{equation}
$A_{T}$ and $A_{P}$ being the target and projectile masses. The
nucleon-nucleon interaction in (\ref{H}) consists of a local
Skyrme interaction, a long-range Yukawa interaction and an
effective charge Coulomb interaction parts \cite{aich, hart}:
\begin{eqnarray}
V_{ij}^{loc}&=& t_{1}\delta({\bf r}_i-{\bf r}_j) +
t_{2}\delta({\bf
r}_i-{\bf r}_j)\delta({\bf r}_i-{\bf r}_k) \label{pot} \\
V_{ij}^{Yuk}&=&t_3 \frac{exp \{ -| {\bf r}_i-{\bf
r}_j|\}/\mu}{|{\bf
r}_i-{\bf r}_j|/\mu} \nonumber\\
V_{ij}^{Coul}&=& \frac {{Z_i}\cdot{Z_j}~e^2}{|{\bf r}_i -{\bf
r}_j|} \nonumber
\end{eqnarray}
$Z_{i}$, $Z_{j}$ are the effective charge of baryons {\it i} and
{\it j}. In QMD model, one neglects the isospin dependence of the
interaction. All nucleons in a nucleus are assigned the effective
charge $Z=\frac{Z_{T}+Z_{P}}{A_{T}+A_{P}}$ \cite{hart}. The
long-range Yukawa force is necessary to improve the surface
properties of the interaction. The parameters $\mu, t_{1}, t_{2},
t_{3}$ in (\ref{pot}) are adjusted and fitted so as to achieve the
correct binding energy and mean square root values of the radius
of the nucleus \cite{aich}. Since QMD model follows the time
evolution of nucleons only, one has to construct the fragments. In
a simplest approach, two nucleons are assumed to share the same
cluster if they are closer than a distance of 4 fm. This method,
also known as minimum spanning tree (MST), can be applied when
matter is dilute and well separated. This picture is true when
incident energy is high and collisions are central in nature. One
has to also keep in the mind that semi-classical models like QMD
can not keep nuclei stable for long time. A typical stability of
nuclei can be seen untill 200 fm/c. If one analyzes  the fragment
formation with MST alone, then one may not achieve true fragment
structure at 200 fm/c. To speed up the recognition of fragment
structure, we add secondary condition that fragments produced with
MST method are subjected to further binding energy check:

\begin{eqnarray}
\zeta &=& \frac{1}{N^{f}}\sum_{i=1}^{N^{f}}\left[\frac {\left(\bf
{p}_{i}-\textbf{P}_{N^{f}}^{cm}\right)^{2}}{2m_{i}}+\frac{1}{2}\sum_{j\neq
i}^{N^{f}}V_{ij}
\left(\bf{r}_{i},\bf{r}_{j}\right)\right]<E_{bind}. \label{bc}
\end{eqnarray}

We take $E_{bind}$ = -4.0 MeV/nucleon if $N^{f}\geq3$ and
$E_{bind} = 0$ otherwise. In this equation, $N^{f}$ is the number
of nucleons in a fragment, $P_{N^{f}}^{cm}$ is the center-of-mass
momentum of the fragment. This modified version of conventional
MST method with binding energy check is labeled as MSTB method.
The magnitude -4.0 MeV/nucleon of $E_{bind}$ is able to recognize
the fragment structure quite accurately. It is chosen keeping in
the mind the average binding energy of clusters. In a recent
communication \cite{jpg}, we used instead microscopic binding
energies based on experimental information. Nearly no effect was
seen by varying the binding energy. We have shown in many
calculations that this check is close to other momentum cuts
\cite{sk98, sk98b, jai, dhe} or sophisticated algorithms like
simulated annealing clusterization algorithm \cite{rkp96, rkp2k,
dh2k7}.

We employ a soft equation of state (EoS) along with Cugnon
parametrization of {\it n-n} cross section for the present study
\cite{aich}. The choice of soft EoS has been advocated in many
theoretical studies. Recently, Magestro \emph{et al}. \cite{mages}
tried to pin down the nuclear incompressibility using balance
energy. Their detailed study pointed towards a \emph{softer}
equation of state. Another study concerning the linear momentum
transfer occuring in central HI collisions \cite{hdad} also showed
that a soft compressibility modulus is needed to explain the
experimental data.

\section{\label{results} Results and Discussion}

Here, we simulate the central heavy-ion collisions of
$^{20}Ne+^{20}Ne$ ($E_{lab}$=10 to 55 AMeV),$^{40}Ar+^{45}Sc$
($E_{lab}$=35 to 115 AMeV), $^{58}Ni+^{58}Ni$ ($E_{lab}$=35 to 95
AMeV) and $^{86}Kr+^{93}Nb$ ($E_{lab}$=35 to 95 AMeV),
$^{129}Xe+^{124}Sn$ ($E_{lab}$=45 to 130 AMeV) and
$^{197}Au+^{197}Au$ ($E_{lab}$=70 to 130 AMeV). The systematic
study over a wide range of beam energies and system masses allows
one to confront the theoretical predictions with experimental
findings and search for the mass dependence. Note that only
symmetric reactions are taken for present analysis. Our
calculations are performed at fixed impact parameter b= 0 fm. We
calculate the reaction at fixed incident energies and then
calculate corresponding center of mass energy. For each such set,
500 events were simulated that minimizes the fluctuations to
greater extent. The choice of central collisions for the present
study guarantees the formation of highly excited systems that may
break into a large number of pieces. Further, the emission from
such events is almost isotropic, which may represent a
`\emph{single source}' emission.
\begin{figure}
\centering
\includegraphics [scale=0.5]{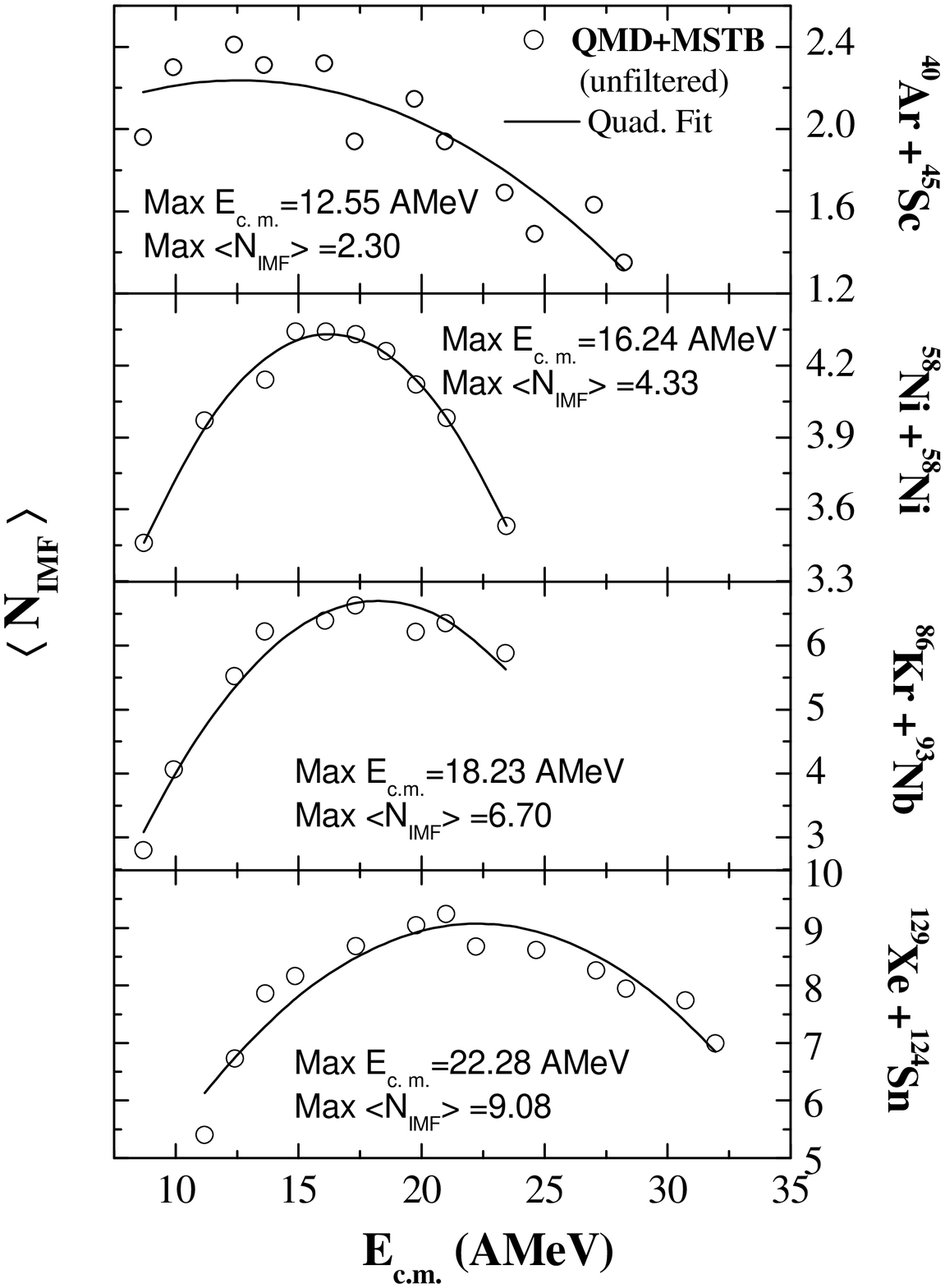}
\vskip -0.33cm \caption{\label{imf}The mean IMF multiplicity
$\langle N_{IMF} \rangle$ versus beam energy $E_{c.m.}$ for the
reaction of $^{40}Ar+^{45}Sc$, $^{58}Ni+^{58}Ni$,
$^{86}Kr+^{93}Nb$ and $^{129}Xe+^{124}Sn$. Open circles depict the
calculations employing QMD + MSTB approach for unfiltered events.
The quadratic fits (solid curves) to the model calculations are
drawn to estimate the peak energy at which the maximal IMF
emission occurs.}
\end{figure}
In figure ~\ref{imf}, we display the average multiplicity of
intermediate mass fragments $\langle N_{IMF} \rangle$ calculated
as a function of beam energy $E_{c.m.}$ in the center-of-mass
frame employing MSTB method. We display here the model
calculations for unfiltered events of four entrance channels
$^{40}Ar+^{45}Sc$, $^{58}Ni+^{58}Ni$, $^{86}Kr+^{93}Nb$ and
$^{129}Xe+^{124}Sn$. The $\langle N_{IMF} \rangle$ first increases
with beam energy, reaches a peak value and then decreases. This
trend is visible in all of the four entrance channels shown here.
This trend is less clear for the lighter $^{40}Ar+^{45}Sc$ system
whereas it is more clearly visible for the heavier systems. A
similar dependence of $\langle N_{IMF} \rangle$ on center of mass
energy is also observed in experimental data taken with the MSU
4$\pi $-Array \cite{sis}. This behavior can be understood in terms
of compression energy of the system. With the rise in the beam
energy, compression energy breaks the IMFs into lighter mass
fragments thereby, leading to fall in the multiplicity of IMFs.
The maximal $E_{c.m.}$ and corresponding peak $\langle N_{IMF}
\rangle$ was obtained through a quadratic fit to the model
calculations. One should also note that the shape of the beam
energy dependence of IMF production is quite close to one reported
in the experimental data \cite{sis}. As reported by Sisan {\it et
al.} \cite{sis}, the peak $E_{c.m.}$ extracted for different
entrance channels scales with the size of the system. Such scaling
is also visible in our present calculations (see Fig.~\ref{imf}).
\begin{figure}[tbp!]
\centering
\includegraphics [scale=0.5]{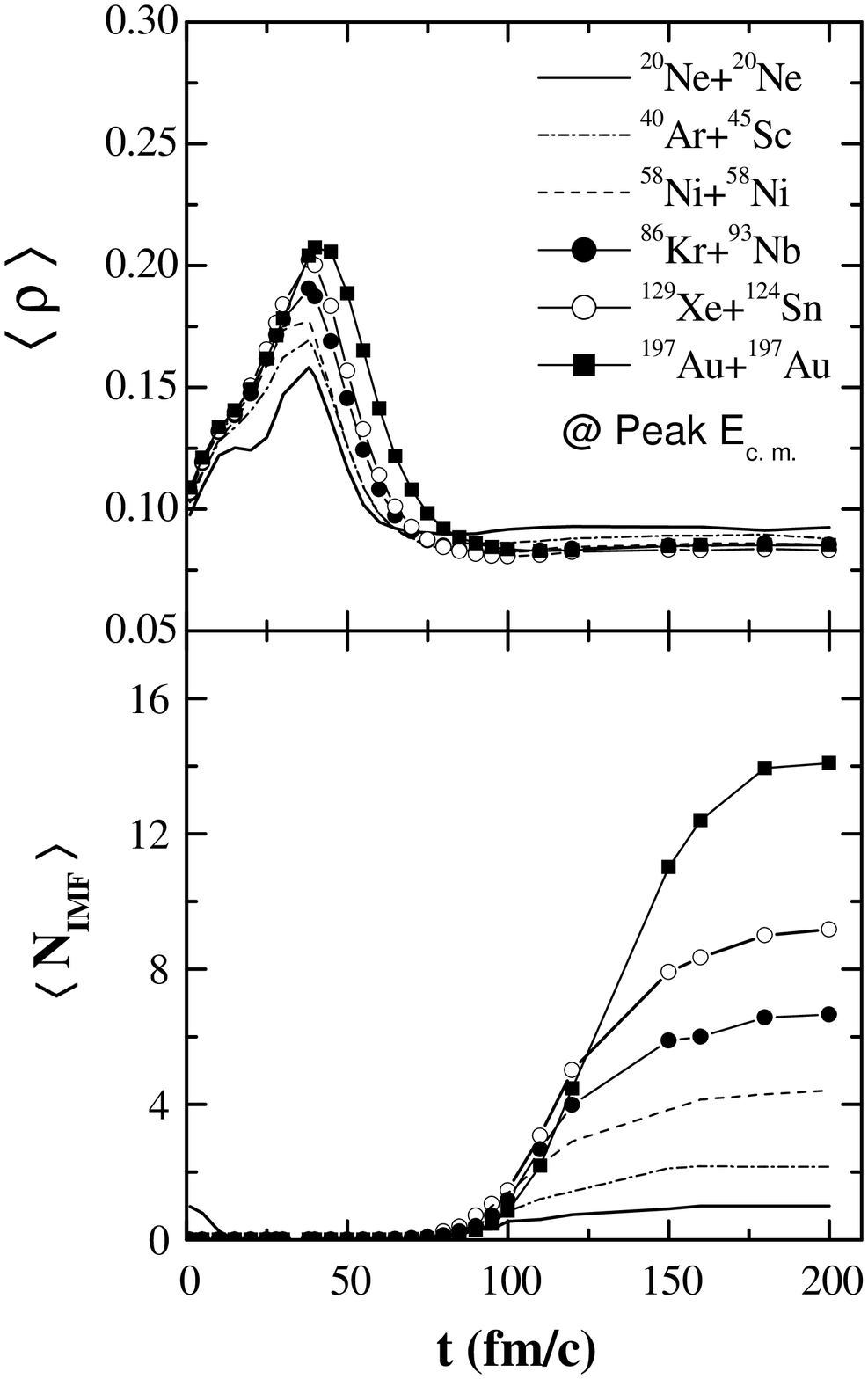}
\vskip -0.52cm \caption{\label{dens} The time evolution of mean
nucleon density (upper panel) and mean IMF multiplicity (lower
panel). Results displayed here are at the energy for peak IMF
production.}
\end{figure}
In figure~\ref{dens}, we display the time evolution of average
density along with the mean multiplicity of intermediate mass
fragments $\langle N_{IMF} \rangle$ defined as fragments $3\leq Z
\leq 20$ plotted at the peak center-of-mass energy $E_{c.m.}$ (at
which maximal IMF emission occurs). We now include
$^{20}Ne+^{20}Ne$ and $^{197}Au+^{197}Au$ systems also for the
study of system size effects. The average nucleonic density of the
system is calculated as:

\begin{equation}
\langle\rho\rangle=\left \langle \frac{1}{N}
\sum_{i=1}^{N}\sum_{j>i}^{N}\frac{1}{(2\pi L)^{3/2}}
e^{-(\bf{r}_{i}(t)-\bf{r}_{j}(t))^{2}/2L} \right \rangle,
\end{equation}

\begin{figure}[tbp!]
\centering
\includegraphics [scale=0.5]{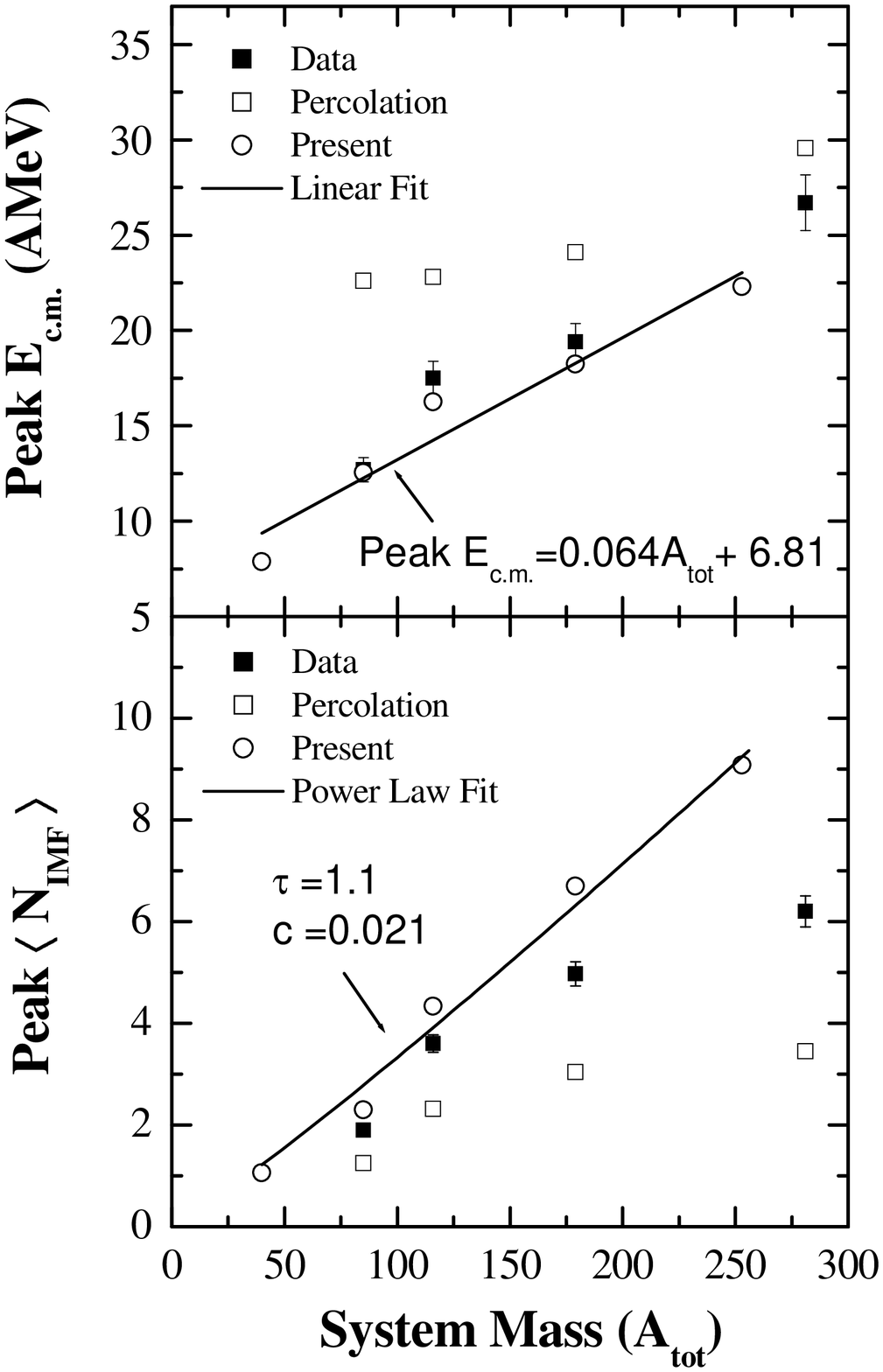}
\vskip -0.52 cm \caption{\label{size} The system size dependence
of the peak $E_{c.m.}$ and peak $\langle N_{IMF} \rangle$. Our
model calculations (open circles) for unfiltered events are
compared with experimental data (solid squares). Also shown in the
figure are the percolation calculations (open squares)
\cite{sis}.}
\end{figure}
with $\mathbf{r}_i$ and $\mathbf{r}_j$ being the position
coordinates of the $i^{th}$ and $j^{th}$ nucleons. The Gaussian
width L is fixed with a standard value of $1.08~fm^{2}$. As
expected, the average nucleonic density has a mass dependence,
being maximal for the $^{197}Au+^{197}Au$ system and minimal for
the $^{20}Ne+^{20}Ne$ system. This also indicates a linear density
dependence on the system size. The intermediate mass fragments
also show similar mass dependence. One can also notice that
fragment production almost saturates around 200 fm/c. In other
words, time span of 200 fm/c is large enough  to pin down the
fragment structure. The maximal fragment production is for
$^{197}Au+^{197}Au$ system whereas $^{20}Ne+^{20}Ne$ system
results in minimum value. It may be mentioned that IMF
multiplicities obtained in $^{20}Ne+^{20}Ne$ and $^{40}Ar+^{45}Sc$
collisions exclude the largest and second largest fragment
respectively to infer the system size dependence accurately.

We plot in figure ~\ref{size}, the peak $E_{c.m.}$ as well as peak
$\langle N_{IMF} \rangle$ as a function of total mass of the
system $A_{tot}$. Strikingly, our model calculations employing
MSTB approach are in good agreement with the experimental data
(solid squares) of  MSU 4$\pi $-Array for peak $E_{c.m.}$. For
peak $\langle N_{IMF} \rangle$, some deviation can be seen for
heavier masses. This could also be due to the fact that our
calculations are not filtered for experimental acceptance.

One can also see that the predictions of percolation model fail to
explain the sharp dependence of peak $E_{c.m.}$ on system mass.
Our present results show a linear mass dependence of the form:
$mA_{tot}+c$ for the peak $E_{c.m.}$. These observations suggest
that the peak $E_{c.m.}$, thus, acts as a measure of \emph{finite
size} effect. It is worth mentioning that the critical excitation
energy was estimated from the cluster size distribution fitted to
power law: $\sigma(A)\propto A^{-\lambda}$ at different beam
energies for which the exponent $\lambda$ reaches a minimum. Based
on the percolation calculations, the critical excitation energy is
also found to increase when initial lattice size increases
\cite{li}. Interestingly, the mass scaling of peak $\langle
N_{IMF} \rangle$ can be reproduced with a power law:
$cA_{tot}^{\tau}$ with exponent close to unity.
\begin{figure}[!t]
\begin{center}
\includegraphics [scale=0.5]{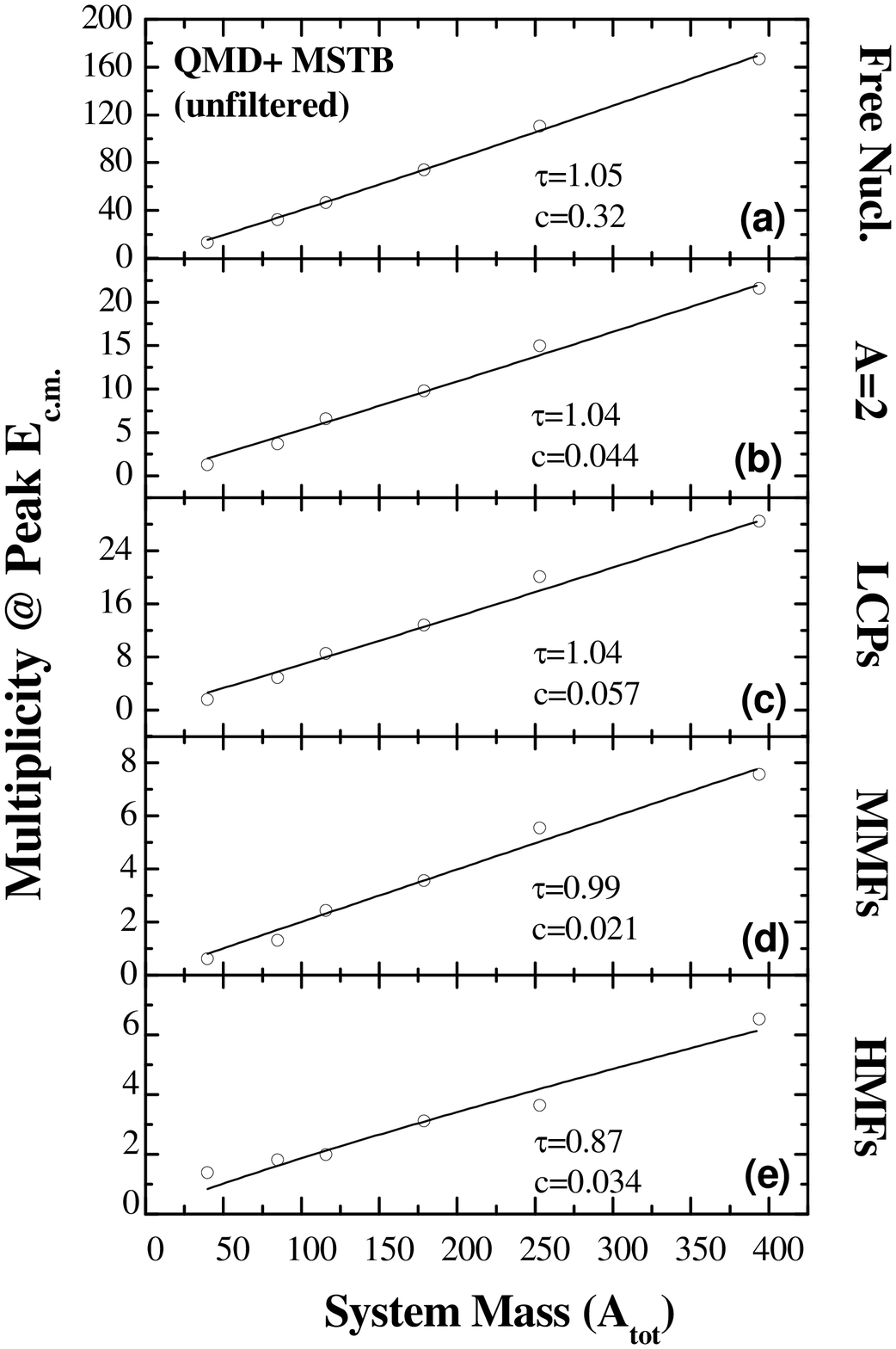}
\vskip -0.64 cm \caption{\label{taus} The multiplicity of (a) free
nucleons, (b) fragments with mass A=2, (c) light charge particles
LCPs, (d) medium mass fragments MMFs, and (e) heavy mass fragments
HMFs as a function of total mass of the system $A_{tot}$. Model
calculations done at peak $E_{c.m.}$ (open circles) are fitted
with power law of the form: $cA_{tot}^{\tau}$.}
\end{center}
\end{figure}
In figure ~\ref{taus}, we finally extend the above study for
various fragments consisting of free nucleons, fragments with mass
A=2, light charged particles LCPs $[2\leq A \leq4]$, medium mass
fragments MMFs $[5 \leq A \leq 9]$ as well as heavy mass fragments
HMFs $[10 \leq A \leq 44]$. Interestingly, in all the above cases,
a clear system size dependence can be seen in a manner similar to
$\langle N_{IMF} \rangle$ dependence. We observe a power law of
the form $cA_{tot}^{\tau}$ ; $A_{tot}$ is mass of the composite.
In all the cases, parameter $\tau$ is very close to unity. As
noted in ~\cite{sis}, the percolation model failed badly to
reproduce the power law dependence. A linear mass dependence
observed with value of $\tau \sim 1$ depicts the picture of
vanishing surface-Coulomb effects. Experiments are called for to
verify this new prediction.

\section{\label{summary} Summary}
We aimed to reveal the dependence of IMF production on beam energy
and system size. This was achieved by a study over wide range of
system masses and incident energies. Our present results reproduce
the experimental trend of both rise and fall in $\langle N_{IMF}
\rangle$ with beam energy. At the point of onset of
multifragmentation, we obtained the scaling of peak $E_{c.m.}$
with system mass. The observed trend of peak center of mass energy
is in agreement with previous experimental studies \cite{li, sis,
stone, Lope, jak, ogi}. The behavior of average nucleon density as
well as IMF multiplicity at the point of onset of
multifragmentation for different entrance channels also reflect
the dominance of system size effects. Our calculations also
reproduce a power law of the form $cA_{tot}^{\tau}$ ; $A_{tot}$
being the total mass of the system. We predict a similar power law
dependence for the fragments of different sizes at the energy for
peak IMF emission.
Interestingly, as observed experimentally, the exponent $\tau$ is close to unity in all cases.\\

This work is supported by CSIR, Government of India vide grant no.
7167/NS-EMR-II/2006, India.


\begin{thebibliography}{999}

\bibitem{leray} Leray S, Ng$\breve{o}$ C, Bouissou P, Remaud B. and
S$\acute{e}$bille F 1991 {\it Nucl .Phys.} A {\bf 531} 177

\bibitem{peilert} Peilert G,  St\"{o}cker H,  Greiner W,  Rosenhauer A, Bohnet A and Aichelin J 1989 {\it Phys. Rev.} C
{\bf 39} 1402

\bibitem{rkp} Puri R K and Kumar S 1998 {\it Phys. Rev.}C {\bf 57} 2744

\bibitem{will} Williams C {\it et al} 1997  {\it Phys. Rev.} C {\bf 55} R2132

\bibitem{blaich} Begemann-Blaich M {\it et al} 1993 {\it Phys. Rev.} C {\bf
48} 610

\bibitem{beau} Beaulieu L {\it et al} 1996 {\it Phys. Rev.} C {\bf 54} R973
\bibitem{li} Li T {\it et al} 1994 {\it Phys. Rev.} C {\bf 49} 1630
\bibitem{li2} Li T {\it et al} 1993 {\it Phys. Rev. Lett.} {\bf 70} 1924
\bibitem{sis} Sisan D {\it et al} 2001 {\it Phys. Rev.} C {\bf 63} 027602
\bibitem{aich} Aichelin J 1985 {\it Phys. Rep.} {\bf 202} 233
\bibitem{hart} Hartnack Ch, Puri R K, Aichelin J, Konopka J, Bass S A, St\"{o}cker H and
Greiner W 1998 {\it Eur. Phys. J. A} {\bf 1} 151
\bibitem{jai2} Singh J and Puri R K 2000 {\it Phys. Rev.}C {\bf 62} 054602
\bibitem{jpg} Vermani Y K, Dhawan J K, Goyal S and Puri R K 2009 {\it J. Phys. G} -communicated

\bibitem{sk98} Kumar S and Puri R K 1998 {\it Phys. Rev.}C {\bf 58} 320
\bibitem{sk98b} Kumar S and Puri R K 1998 {\it Phys. Rev.}C {\bf 58} 2858
\bibitem{jai} Singh J and Puri R K 2001 {\it J. Phys. G}{\bf 27} 2091
\bibitem{dhe}Dhawan J K and Puri R K 2007 {\it Eur. Phys. J. A} {\bf 33} 57

\bibitem{rkp96} Puri R K, Hartnack C and  Aichelin J 1996 {\it Phys. Rev.}{\bf 54}
R28; Nebauer R, Guertin A,  Puri R K, Hartnack C, Gossiaux P B and
Aichelin J 1999 {\it Proc. Int. Wrks. on Gross Properties of
Nuclei and Nuclear Excitations (Hirschegg Austria)}
Vol~\textbf{27} ed by H. Feldmeier {\it et al} (Darmstadt:GSI)
p~43
\bibitem{rkp2k} Puri R K and Aichelin J 2000 {\it J. Comput. Phys.} {\bf 162} 245
\bibitem{dh2k7} Dhawan J K and Puri R K 2007 {\it Phys. Rev.} C {\bf 75} 057601

\bibitem{mages} Magestro D J,  Bauer W and Westfall G D 2000 {\it Phys. Rev.} C {\bf
62} 041603(R)

\bibitem{hdad} Haddad F {\it et al} 1996 {\it Phys. Rev.}C {\bf 53} 1437
\bibitem{stone} Stone N T B, Llope W J and Westfall G D 1995 {\it Phys. Rev.} C {\bf 51} 3157
\bibitem{Lope} Llope W J {\it et al} 1995 {\it Phys. Rev.}C {\bf 51} 1325
\bibitem{jak} Jakobsson B {\it et al} 1990 {\it Nucl .Phys.}A {\bf 509} 195
\bibitem{ogi} Ogilvie C A {\it et al} 1991 {\it Phys. Rev. Lett.}{\bf 67} 1214

\end{thebibliography}
\end{document}